\documentclass[pra,twocolumn,showpacs]{revtex4}
\usepackage{bm,amsmath,amsfonts,amsbsy}
\usepackage{graphicx}
\def \d3k {\frac{d^3k}{(2\pi)^3}}
\def \v#1{{\bm #1}}

\def \be {\begin{equation}}
\def \ee {\end{equation}}

\begin{document}

\title{Critical Velocities for Energy Dissipation from Periodic Motions of Impurity in Bose-Einstein Condensates}
\author{Jun Suzuki}
\email{physj@nus.edu.sg}
\affiliation{Department of Physics, National University of Singapore\\
Singapore 117542, Singapore}
\date{\today}
\begin{abstract}
A phenomenon of energy dissipation in Bose-Einstein condensates is studied based on 
a microscopic model for the motion of impurity. Critical velocities for onset of energy dissipation 
are obtained for periodic motions, such as a dipole-like oscillation and a circular motion. 
The first example is similar to a series of MIT group experiments settings where the critical velocity 
was observed much below the Landau critical velocity.  The appearance of the smaller 
values for the critical velocity is also found in our model, even in the homogeneous 
condensate in the thermodynamic limit.  This suggests that the landau criterion 
be reexamined in the absence of quantized vortices in the bulk limit. 
\end{abstract}
\maketitle

%%%%%%%%%%%%%%%%%%%%%%%%%%%
\section{Introduction}
%%%%%%%%%%%%%%%%%%%%%%%%%%%
In 1941 Landau gave a phenomenological argument about the critical 
velocity in a superfluid below which no energy dissipations occur \cite{landau}. 
His argument was entirely based on a kinematics and the 
Galilean invariance. This critical velocity is now known as the
Landau criterion, which states that if impurities in a superfluid
move slower than $v_c=\min \omega(\v p)/|\v p|$, then there are no
excitations created in a superfluid. Here $\omega(\v p)$ is the
dispersion relation of excitations with the momentum $\v p$. This
then explains the phenomenon of the superfluidity. In a real
superfluid, however, finite energy dissipations take place even
below the Landau critical velocity. This is due to creation of
quantized vortices, which was explained later by Feynman \cite{feynman}. 
A microscopic understanding of the superfluidity was
initiated by Bogoliubov \cite{bogoliubov47}. He particularly showed that 
the phenomenon of superfluidity can be explained as a consequence of 
condensation of massive bosons in the ground state. Although his model
cannot be strictly speaking, applied to the real $^4$He due to
strong interactions between atoms, he succeed to explain the Landau 
criterion from the first principle. Bogoliubov's weakly interacting massive 
boson model has been studied extensively in more than half century \cite{bru}.

The celebrated experimental realization of Bose-Einstein condensates
(BECs) in several alkali vapors have revived theoretical works on
BECs as well as experimental works \cite{bec,review}. Particularly, the 
mean field approximation of the Bogoliubov model, the Gross-Pitaevskii equation, 
has been studied extensively \cite{review}. 
One reason is because this model can be adopted as a realistic model.
Another reason is that the current experimental technics are
remarkably well developed and under controlled to provide variety of opportunities to test
these theories. Indeed, so far theories have good
agreements with experimental results in many cases. 
However, there are still experimental facts which are left without
satisfactory explanation. In this paper we would like to give 
analytical argument to one of those experiments. 
They are a series of experiments done by MIT group \cite{mit1,mit2,mit3}.

These experiment were intended to examine the Landau criterion in
BECs. Interestingly, they have found disagreement with Landau's argument. 
The essence of these experiments is as follow. The sodium
condensate was created in anisotropic harmonic traps forming an elongated in one 
axial direction. A Gaussian laser beam was then used to stir the
condensate by moving it back and forth periodically. The critical
velocity for creating energy dissipation was observed less ten times
smaller than the Landau critical velocity. Since a publication of
these experimental results, there appear many theoretical works to
fulfill this discrepancy. One of the important question is whether
this experimental fact is due to the bulk property of BECs or is
originated from other factors, such as creation of vortices
\cite{jackson}, the geometry of the condensate \cite{fedichev}, and
so on \cite{ref}. There is no doubt about the fact that all these
factors give rise to the experimental observations of energy
dissipations below the Landau critical velocity. However, there still remain 
unclear whether this discrepancy may happen in the homogeneous system. 

In this paper we argue that the appearance of the smaller critical
velocity is also part of the bulk properties of the condensates.
Hence this effect does occur even in the homogeneous system in the
thermodynamic limit as contrast to the original Landau's argument. 
To demonstrate it we will evaluate the critical velocity for the energy dissipation 
in the homogeneous BEC in the thermodynamics limit at zero temperature. 
To capture the physics of MIT experiments, we 
couple the weakly interacting bosons with an external impurity whose
trajectory is given by a periodic motion similar to the actual
experiment. We show that this model provides qualitative
understandings of the discrepancy observed in laboratories. 
We also compare the result with other possible periodic motion, a circular motion. 
The result shows that the same conclusion holds for the case of the circular motion. 
A comparison between two cases suggests that the circular motion has more dissipation 
than the dipole-like oscillation and the larger critical velocity. 

The paper continues as follows. 
In Sec.~II we give a summary of our model and its solution within the Bogoliubov approximation.  
A formula for the energy dissipation due to the motion of impurity in BECs is also given. 
An idealized case of MIT group experimental settings is studied and the critical velocity is evaluated in Sec.~III. 
In Sec.~IV energy dissipation from a circular motion and its comparison to the result of Sec.~III are given. 
Sec.~V gives a summary and discussion of our results.

%%%%%%%%%%%%%%%%%%%%%%%%%%%%%%%%%
\section{The Model and Its Solution}
%%%%%%%%%%%%%%%%%%%%%%%%%%%%%%%%%

\subsection{The model Hamiltonian and its diagonalization}
The model for motions of classical impurities in the homogeneous BEC
was proposed and studied in \cite{ms,jun1,jun2}.  We give a summary of
result together with the essence of the model. The total Hamiltonian
is the sum of two parts. One is the standard Bogoliubov weakly
interacting massive bosons term $\hat{H}_B$, and the other is the
interaction term $\hat{H}_I$ between bosons and a moving impurity
which takes into account the local interaction between them : 
\begin{multline}
\hat{H}_B= \int\!d^3x\;\hat{\psi}^{\dagger} (\v{ x},t) (-\frac{\hbar
^2 \v{ \nabla} ^2}{2 M})\hat{\psi} (\v{ x},t) \\+ \frac g2
\int\!d^3x\;\hat{\psi}^{\dagger} (\v{ x},t) \hat{\psi}^{\dagger}
(\v{x},t)\hat{\psi} (\v{ x},t) \hat{\psi} (\v{x},t) , 
\end{multline}
where $M$ is the mass of the bosons. The coupling constant $g$ between bosons
is expressed in terms of the s-wave scattering length $a_s$; $g=4
\pi a_s \hbar ^2/M$. We assume the repulsive interaction $g>0$ and
diluteness $na_s^3\ll1$ with $n$ the number density of bosons. The
interaction term is 
\be 
\hat{H}_{I} = \lambda  \int d^3x\;\rho _c
(\v{ x},t) \hat{\psi}^{\dagger} (\v{ x},t) \hat{\psi} (\v{ x},t) .
\ee 
Here $\lambda$ is the coupling constant between bosons and the impurity, 
and $\rho_c (\v{ x},t)$ is the time dependent distribution of classical impurity.
Particularly, we have in mind an idealized point-like disturbance on
the condensate guided along a given trajectory $\v{ \zeta}(t)$, i.e.
we will adopt the replacement $\rho_c (\v{ x},t)=\delta (\v x -\v{\zeta}(t))$.

In order to estimate the effects of the impurity in the thermodynamics limit,
we expand the field operators $\hat{\psi} (\v{ x},t) $ in terms of the plane wave basis with
periodic boundary conditions in a finite size box $V=L^3$. The limit $N$ and $V$ go to
infinity whit a fixed density $n=N/V$ will be taken at the end of calculations.
We follow Bogoliubov's treatment to simplify the total Hamiltonian $\hat{H}=\hat{H}_B +\hat{H}_{I}$
within a number conserving framework of Bogoliubov model \cite{comment}.
After several steps \cite{jun2}, we arrived at an approximated Hamiltonian :
\begin{multline}
\hat{H}_B  =E_0 + \sideset{}{'}\sum (\epsilon _{k}+gn) \hat{\alpha}_{\v k}^{\dagger}\hat{\alpha}_{\v k}
+ \frac{gn}{2} \sideset{}{'}\sum ( \hat{\alpha}_{\v k}^{\dagger}\hat{\alpha}^{\dagger}_{-\v k} + {\rm h.c.})\\
+ \frac {n \lambda}{\sqrt{N}} \sideset{}{'}\sum (\tilde{\rho} _{\v k}(t) \hat{\alpha}_{\v k}^{\dagger} +{\rm h.c.}) .
\end{multline}
Here $E_0 = g n N/2 -gn/2+ \lambda n$ is a constant term,
$ \epsilon _{k}=\hbar ^2 \v k^2/2M$ is the free kinetic energy of bosons,
and $\tilde{\rho} _{\v k}(t) =\int d^3 x \;\rho_c(\v x, t)e^{-i\v k \cdot \v x}$.
In this approximation we neglected terms of order of $N^{-1}$.

Bogoliubov's excitation is created and annihilated by the operators
\begin{align}
\hat{b}_{\v k}^{\dagger} &= \hat{\alpha}_{\v k} ^{\dagger} \cosh \theta _{k} +\hat{\alpha}_{-\v k} \sinh \theta _{k}, \\
\hat{b}_{\v k}&=\hat{\alpha}_{\v k}  \cosh \theta _{k} + \hat{\alpha}_{-\v k}^{\dagger} \sinh \theta _{k},
\end{align}
with
\be
 \theta _{k} = \tanh ^{-1}(\frac{gn}{\hbar \omega _{k} +\epsilon _{k}+gn}), \ \hbar \omega _{k}
 = \sqrt{\epsilon _{k}(\epsilon _{k}+2 gn)} .
\ee
In terms of Bogoliubov's excitation, the total Hamiltonian is expressed as
\be \label{ha2}
\hat{H}= E'_0
+\sideset{}{'} \sum \hbar \omega _{k} \hat{b}_{\v k}^{\dagger}  \hat{b}_{\v k}
+ \sideset{}{'} \sum \v (f_{\v k}(t) \hat{b}_{\v k}^{\dagger}+{\rm h.c.}\v ) .
\ee
Here $E'_0= E_0 + \sideset{}{'} \sum  (\hbar \omega _{k} -\epsilon _{k}-gn)/2$
is the ground state energy without impurities (a prime indicates the omission of the zero mode from the summation),
and $f_{\v k}(t)= n \lambda \tilde{\rho} _{\v k}(t) \sqrt{\epsilon _{k}/(N \hbar \omega _{k})}$.
Therefore, the fist order impurity effects are equivalent to decoupled forced harmonic oscillators,
which can be solved analytically. Since we are solving the time dependent problem,
it is natural to work in the Heisenberg picture. The equations of motions for Bogoliubov's excitation
creation and annihilation operators are easily solved as follows.
\begin{align} \label{solution1}
\hat{b}^{\dagger}_{\v k}(t)&=\hat{B}^{\dagger}_{\v k}(t)+\phi ^*_{\v k}(t),\\
\hat{b}_{\v k}(t)&=\hat{B}_{\v k}(t)+\phi _{\v k}(t) .\label{solution2}
\end{align}
Here $\phi _{\v k}(t)$ is a $c$-number function :
\be  \label{phi}
\phi _{\v k}(t) = \frac{n\lambda}{i\hbar} \sqrt{\frac{\epsilon _{k}}{N \hbar \omega _{k}}}I_{\v k} (t)e^{-i \omega _{k} t}  ,
\ee
where the integral $I_{\v k}(t)$ is defined by
\be \label{integral}
I_{\v k}(t) =\int ^t _{t_0} d t'\;\tilde{\rho}_{\v k} (t')e^{i\omega_k t'}.
\ee
We have chosen the boundary condition such that the system is disturbed by impurity at $t=t_0$.

Now the our system is described by in terms of the dressed
Bogoliubov excitation creation and annihilation operators
$\hat{B}^{\dagger}_{\v k}$ and $\hat{B}_{\v k}$. They evolve
according to the diagonalized Hamiltonian $\hat{H}' $ : \be \hat{H}'
= \tilde{E}_0(t)+ \sideset{}{'} \sum \hbar \omega _{k} \hat{B}^{
\dagger}_{\v k} \hat{B}_{\v k}. \ee In the diagonalized Hamiltonian,
the new ground state energy is defined by $\tilde{E}_0(t)= E'_0+
\sideset{}{'} \sum {\rm Re}\v (f_{\v k}^*(t)\phi _{\v k}(t)\v )$.
The energy spectrum $\hbar \omega _{k}$ is that of the gapless
excitations characterized by $\omega _{k} \simeq kc_s$ for a small
$k$, where $c_s=\sqrt{gn/M}$ is the speed of sound. We remark that
the spectrum $\omega_k$ and the speed of sound $c$ are the same as
in the original Bogoliubov model without impurities. Hence, 
the motion of impurities does not affect either $\omega_k$ nor $c$
in our model within the above approximation.

\subsection{Homogeneous condensate}
The homogeneous BEC is defined by the ground state of the Hamiltonian (\ref{ha2}) without the impurity term.
In other word the vacuum state of the annihilation operators $\hat{b}_{\v k}$.
We denote it as $|bec\rangle$, and this is to be defined by
\be
\hat{b}_{\v k}|bec\rangle=0\quad \forall \v k \neq 0.
\ee
Therefore, we conclude that the dressed Bogoliubov excitation annihilation operator due to impurity
{\it does not} see the condensate as vacuum :
\be \label{result}
\hat{B}_{\v k} |bec\rangle =- \phi _{\v k}(t) |bec\rangle .
\ee
Alternatively, one can interpret this result as follows.  
A classical field describe by $\phi _{\v k}(t) $ is induced in the homogeneous condensate 
due to the motion of impurity. 

\subsection{Energy dissipation due to the motion of impurity}
As we have mentioned in Introduction, we are interested in the
amount of dissipated energy in condensates due to the motion of
impurity. Physically, energy dissipations take place via scattering
processes between bosons and impurity. In other words the impurity
motion creates excitations in condensates by scattering processes.
Excitations in BECs will carry amount of energy equal to their
excitation spectra.  The amount of dissipated energy is then
measured by heat transferred to condensates.

We first define the occupation number for the dressed Bogoliubov's excitation with respect to the homogeneous condensate.
This number counts the emitted Bogoliubov's excitations accompanying with the motion of impurity in BEC,
\be \label{occu}
\tilde{n}_{\v k}(t) \equiv \langle bec| \hat{B}^{\dagger}_{\v k} (t)\hat{B}_{\v k} (t) |bec \rangle
= |\phi _{\v k} (t)|^2% =\frac{n^2 \lambda ^2 \epsilon_k}{N \hbar^3 \omega_{k}}|I_{\v k}(t)|^2.
\ee
Multiplying by the excitation energy $\hbar \omega_{k}$ gives the dissipated energy
${\cal E}_{\v k} (t)$ for a given mode $\v k$, and the total dissipated energy ${\cal E} (t)$ is given by summing over all modes.
In our model we assume that an impurity is driven by some external forces, in which
back reaction is negligible. Therefore, this dissipated energy is equivalent to
the amount of energy transfered to the condensate from the external impurity.
In the thermodynamic limit, the total energy dissipation due to impurity is evaluated
by rather simple formula :
\be \label{diss}
{\cal E} (t)= \sideset{}{'} \sum {\cal E}_{\v k} (t) \to
\frac{n \lambda ^2 }{ \hbar^2} \int\!\d3k \;\epsilon_k |I_{\v k}(t)|^2.
\ee

\subsection{Depletion of the condensate}
The depletion of the condensate $d(t)$ due to the quantum fluctuation is defined by
the formula :
\be
d(t)=\sideset{}{'} \sum\langle bec|\hat{\alpha}_{\v k}^{\dagger}\hat{\alpha}_{\v k}|bec\rangle /N.
\ee
Using the Bogoliubov transformation and the solution obtained before,
\begin{multline}
d(t)= \sideset{}{'} \sum \frac {1}{2N}( \frac{\epsilon _{k} +gn}{\hbar \omega _{k}}-1)\\
+ \frac 1N  \sideset{}{'} \sum (\frac{\epsilon _{k}}{\hbar \omega _{k}}|\phi _{\v k}(t)|^2
+\frac{gn}{2\hbar \omega _{k}}|\phi^*_{\v k}(t)-\phi_{-\v k}(t)|^2) .
\end{multline}
Since $|\phi_{\v k}(t)|^2$ has an additional factor $1/N$ (see eq.~(\ref{phi})), 
the motion of impurity does not contribute
to the depletion of the homogeneous condensate in the thermodynamic limit.
The depletion is then given by $d= (8/3) \sqrt{n a_s^3/\pi}$ which is independent of the impurity effects.
Therefore, in the large $N$ limit, the homogeneous condensate is stable against
an external impurity.
In contrast, however, these terms in the depletion of condensates is not negligible 
in real experiments where the finiteness of the number of particles and finite size effects play important roles.

%%%%%%%%%%%%%%%%%%%%%%%%%%%%%%%
\section{MIT Group Experiments}
%%%%%%%%%%%%%%%%%%%%%%%%%%%%%%%
We now apply our model to the experimental settings done by MIT
group. We consider a point-like impurity which is oscillating along
the $z$-axis with a period $2\pi/\Omega$ and the oscillation
amplitude $d$. The trajectory is expressed by $\v{\zeta}(t)=(0,0,d \cos\Omega t) $.
Although this is still an idealized model, we will see that this
model gives qualitative explanation for the real experiments. To
make the corresponding to the experiments, we also define the
velocity of the impurity by $v=2d/\pi \Omega$. In order to obtain
analytical expressions for the energy transferred to the condensate,
we further make an assumption that the impurity is oscillating for a
sufficiently very long time. This assumption seems reasonable if we
look the real experimental values. Under this assumption we let the
time integral (\ref{integral}) from $-\infty$ to $+\infty$. Using
the plane wave expansion in terms of the Bessel function ${\rm J}_{\ell}(x)$, 
$\exp(-ix\cos\phi)=\sum_{\ell=-\infty}^{\infty}(-i)^{\ell}\exp(i\ell\phi){\rm J}_{\ell}(x)$, 
we obtain
\begin{align}
I_{\v k}(\infty)
&=\sum_{\ell=-\infty}^{\infty}(-i)^{\ell}{\rm J}_{\ell}(kd\cos\theta) 
\int_{-\infty}^{\infty}\!dt\;e^{i\omega_k t+i \ell \Omega t}\\
&=2\pi\sum_{\ell=1}^{\infty}(-i)^{\ell}{\rm J}_{\ell}(kd\cos\theta)\;\delta(\omega_k -\ell\Omega). 
\end{align}
Here the wave number vector $\v k$ is written in the spherical
coordinates with $\theta$ an angle from the $z$-axis. Therefore, in
the idealized infinite limit, the integral in consideration will
take discrete values labeled by the integer $\ell$. In this limit an
additional care has to be take to evaluate the dissipated energy
since we cannot take the square of the Dirac delta distributions. To
this end we first evaluate the finite time interval $T$, then we
take the limit $T\to \infty$ in the last step. After straightforward but tedious computations, 
we find the total dissipated energy per unit time is
\begin{align} 
\Gamma &\equiv\lim _{T\to\infty} \frac{{\cal E}_{{\rm tot}}(T)}{T}\\
&= \frac{2\pi n \lambda ^2}{\hbar^2}\sum_{\ell=1}^{\infty}
 \int\!\d3k \;\epsilon_k \v ({\rm J}_{\ell}(kd\cos\theta)\v )^2\;\delta(\omega_k -\ell\Omega).
\end{align} 
The above integrals can be evaluated in terms of the
general hypergeometric function : 
\be
_{2}{\rm F}_{3}(a,b;c,d,e;z)=\sum_{n=0}^{\infty}\frac{(a)_n (b)_n}{(c)_n(d)_n (e)_n}\;\frac{z^n}{n!} , 
\ee 
where $(a)_n=\Gamma(a+n)/\Gamma(a)$ with $\Gamma(x)$ the Gamma function.
The final expression for energy dissipation is 
\begin{multline}
\Gamma=\Gamma_0 \sum_{\ell=1}^{\infty} \frac{(k_{\ell}d/2)^{2 \ell}}{(2 \ell+1)(\ell !)^2}
\; \frac{\ell(k_{\ell}\xi)^3}{1+2 (k_{\ell}\xi)^2} \\
\times_{2}{\rm F}_{3}(\ell+\frac 12,\ell+\frac12;\ell+1,\ell+\frac 32 , 2 \ell+1 ;-(k_{\ell}d)^2). 
\end{multline}
Here
$\xi=\hbar/(2 M c_s)$ is the healing length of the condensate, and
$\Gamma_0$ is a constant with dimension of energy per unit time ;
$\Gamma_0=4n\lambda^2M^2c_s\Omega/(\pi\hbar^3)$. The wave number
takes the discrete values as a consequence of the Dirac delta
distribution as 
\be \label{kell} 
k_{\ell}=\frac{1}{\sqrt 2 \xi}\left[\sqrt{1+(\ell\xi\Omega/c_s)^2}-1\right]^{1/2} \quad (\ell =1, 2, \cdots). 
\ee

We plot the dissipated energy per unit time as a function of the
velocity of impurity $v=2d\Omega/\pi$. The velocity is in units of the speed of sound $c_s$, 
and the energy per unit time is scaled by its value for $v=c_s$. 
Several curves correspond to various values of the oscillation amplitude $d$ in units of $\xi$; 
$d/\xi=1,10,100$, and $1000$. Except for the case $d/\xi=1$, all curves merge to the same curve. 
This universal behavior shows that critical velocities are the characteristics of 
the motion of impurity itself and are independent of the parameters involved. 
This statement seems true in general in the bulk limit of condensates. 
\begin{figure}[htbp]
   \centering
   \includegraphics[width=3in]{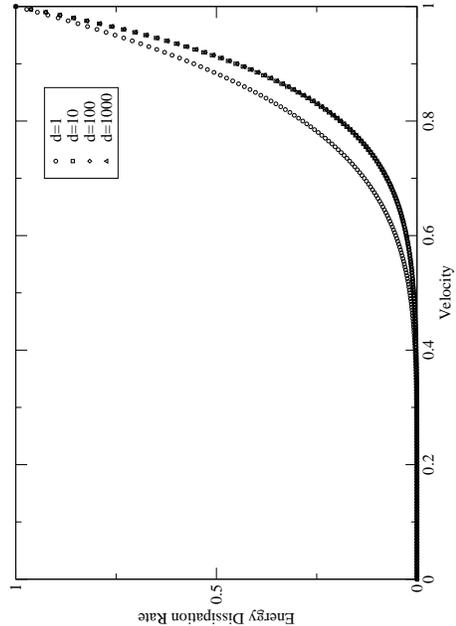}
   \caption{The dissipated energy per unit time for the dipole-like oscillation 
   as a function of the velocity $v$. The velocity is plotted in units of the speed of sound $c_s$, 
   and the energy per unit time is scaled by its value for $v=c_s$. There are four curves which correspond 
   to the values of amplitude; $d/\xi=1,10,100,1000$.}
   \label{fig1}
\end{figure}

We estimate the value for critical velocity $v_c$ by fitting the curve with the expression 
$\Gamma\propto v(v-v_c)$ for small values of $v$. 
This is based on the standard argument that the condensates will experience the drag force 
from the impurity, which is proportional to $\v{v}-\v{v_c}$. The inner product of this force 
and the velocity of impurity then gives the rate of dissipated energy in the condensates.  
The estimated value of the critical velocity for the case $d/\xi=100$ is found as $v_c= 0.34\;c_s$.

%%%%%%%%%%%%%%%%%%%%%%%%%%%%%%%
\section{Energy Dissipation from a Circular Motion}
%%%%%%%%%%%%%%%%%%%%%%%%%%%%%%%
As an another example for periodic motions, we consider a circular
motion on the $xy$-plane, which was studied in details in \cite{jun2}. 
The trajectory is specified by two parameters, the radius $R$ and the angular velocity $\Omega$; 
$\v{\zeta}(t)=(R \cos \Omega t,R \sin \Omega t,0) .$
Following the same calculation methods carried out for the previous section, 
the formula for the dissipated energy per unit time from the circular motion is 
\be 
\Gamma=\sum_{\ell=1}^{\infty}\Gamma_0 \frac{\ell (k_{\ell}\xi)^2}{1+(k_{\ell}\xi)^2}
\sum_{j=0}^{\infty}{\rm J}_{2j+2\ell+1}(2k_{\ell}R), 
\ee 
where $\Gamma_0=n\lambda^2M\Omega/(2\pi\hbar^2R)$, and the discrete wave
number $k_{\ell}$ is given by eq. (\ref{kell}). 
Note that in the circular motion case, the velocity of the impurity is given by $v=R\Omega$ instead.

\begin{figure}[htbp]
   \centering
   \includegraphics[width=3in]{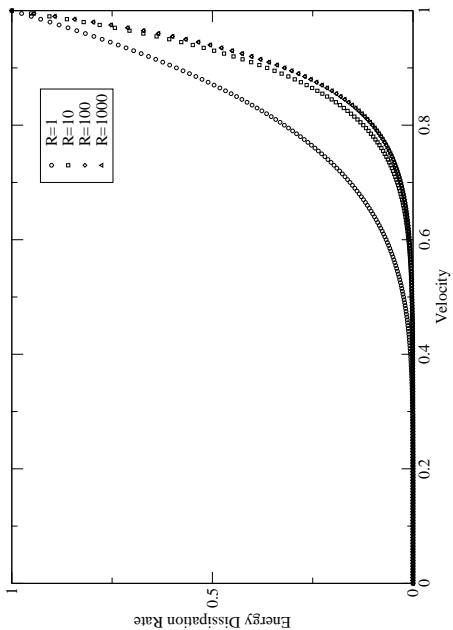}
   \caption{The dissipated energy per unit time for the circular motion as a function of the velocity $v$. 
   The same convention for the units of energy per unit time and the velocity is used as in FIG.~\ref{fig1}.}
   \label{fig2}
\end{figure}

We plot the the dissipated energy per unit time in FIG.~\ref{fig2} in the same manner as FIG.~\ref{fig1}.  
The critical velocity for the circular motion is estimated from the graph as $v_c=0.43\;c_s$. 
This value is slightly bigger than the case of the dipole oscillation. 

\begin{figure}[htbp]
   \centering
   \includegraphics[width=3in]{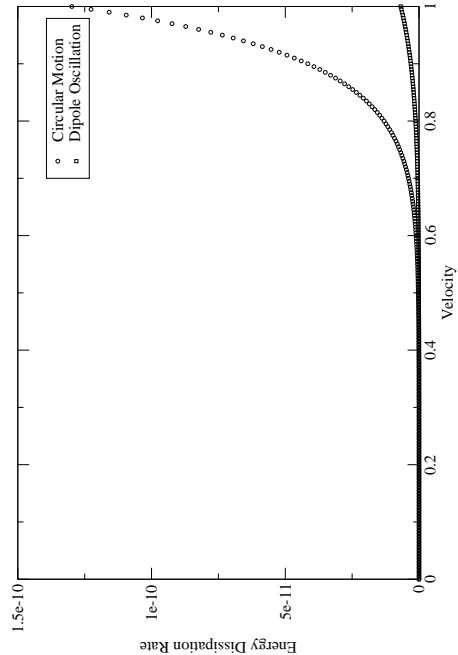}
   \caption{The dissipated energy per unit time for two examples for the amplitude $d/\xi=100$ 
   and the rotation radius $R/\xi=100$.  
   The energy per unit time is in units of $\gamma_0=n\lambda^2/(2\hbar\xi^3)$.}
   \label{fig3}
\end{figure}

For a comparison, we also plot two cases in FIG.~\ref{fig3} for the value $d/\xi=100$ and $R/\xi=100$. 
The units of the dissipated energy per unit time in FIG.~\ref{fig3} is $\gamma_0=n\lambda^2/(2\hbar\xi^3)$.
Two curves show the similarity between two cases. 
Although we cannot make one to one corresponding for the velocities of two cases, 
the circular motion has more energy dissipation in general.   

%%%%%%%%%%%%%%%%%%%%%%%%%%%%%%%
\section{Conclusions and Outlook}
%%%%%%%%%%%%%%%%%%%%%%%%%%%%%%%
We have shown that finite energy dissipation from the periodic motions of impurity can take place even below 
the Landau critical velocity in the homogeneous BEC. We have found the critical velocities $v_c$ 
for the dipole-like oscillation and the circular motion as $0.34;c_s$ and $0.43\;c_s$ respectively. 
Our result qualitatively agrees with the experimental observations of Ref.~\cite{mit1,mit2,mit3}. 
The differences arise from the fact that there are many other factors as pointed out in previous studies 
\cite{jackson,fedichev,ref}, as well as several simplifications in our model and calculations. 
It is, therefore, necessary to extend to our model to the case for BECs with trapping 
potentials at finite temperature to see whether our model yields better agreement or not. 

The result present suggest that the landau criterion be reexamined even 
in the absence of quantized vortices for the homogeneous system.  
A simple physical argument is that the original Landau's argument cannot 
be applied to the case where impurities move under the acceleration.  
This is because the Galilean invariance fails if there is an acceleration between two coordinates systems. 
As a related issue, the phenomenon of Cherenkov-like radiation in BECs was studied 
for an impurity moving with a constant velocity, where the critical velocity for a radiation was found 
to be exactly same as the Landau critical velocity \cite{kovrizhin,astracharchik,jun1,jun2}. 

In this paper we also evaluate the dissipated energy from the circular motion of impurity. 
A trajectory of the circular motion in BECs is in principle experimentally realizable in current experimental technics. 
The studies of impurities under the circular motion in BECs may give us further understanding 
of the coherent properties of condensates and the phenomenon of superfluidities in BECs.

\begin{acknowledgments}
This work was supported in part by the National University of Singapore. 
\end{acknowledgments}  

%%%%%%%%%%%%%%%%%%%%%%%%%%%%%%%%%%%%%%%%%%%


\begin{thebibliography}{99}
\bibitem{ref} There have been many theoretical works related MIT experiments, 
see for instance \cite{potting,montina,mazets,astracharchik,pomeau} and references therein.
\bibitem{comment} There are several proposals for the number conserving treatment for BEC 
\cite{girardeau, gardiner, castin}.  We follow the one used in refs.~\cite{girardeau}.
\end{thebibliography}
\end{document}